# Limitations of Curl and Directional Filters in Elastography

**Kevin J. Parker [1],\***

[1] Department of Electrical and Computer Engineering, University of Rochester, Rochester, NY 14627, USA; e-kevin.parker@rochester.edu

\* Correspondence: kevin.parker@rochester.edu

**Abstract:** In the approaches to elastography, two mathematical operations have been frequently applied to improve the final estimate of shear wave speed and shear modulus of tissues. The vector curl operator can separate out the transverse component of a complicated displacement field, and directional filters can separate distinct orientations of wave propagation. However, there are practical limitations that can prevent the intended improvement in elastography estimates. Some simple configurations of wavefields relevant to elastography are examined against theoretical models within the semi-infinite elastic medium and guided waves in a bounded medium. The Miller-Pursey solutions in simplified form are examined for the semi-infinite medium and the Lamb wave symmetric form is considered for the guided wave structure. In both cases, wave combinations along with practical limits on the imaging plane can prevent the curl and directional filter operations from directly providing an improved measure of shear wave speed and shear modulus. Additional limits on signal-to-noise and the support of filters also restrict the applicability of these strategies for improving elastographic measures. Practical implementations of shear wave excitations applied to the body and to bounded structures within the body can involve waves that are not easily resolved by the vector curl operator and directional filters. These limits may be overcome by more advanced strategies or simple improvements in baseline parameters including the size of the region of interest and the number of shear waves propagated within.

## 1. Introduction

The use of the vector curl operator and directional filters have been key elements in a number of elastography techniques. The curl is used because shear waves can be separated from any irrotational (dilatational or longitudinal) waves by the curl operator [1, 2]. The curl operator can only be applied rigorously to the components of displacement vector fields in three dimensions (3D), and so has been used in conjunction with 3D magnetic resonance imaging (MRI) data sets [3-7], and in fewer ultrasound systems [8, 9].

Directional filters are used where there is a desire to separate out shear wave propagation from different quadrants or angles, so as to capture separate estimations of their behavior. These have been utilized in ultrasound, optical, and MRI applications [10-17].

Given the well-established use of these processes in elastography, it may seem that they will generally lead to an improved estimate of shear wave speed. However, there are limits to both of these approaches and simple configurations where neither the curl operator nor the directional filters lead directly to a more accurate estimate of shear modulus. These practical limits are explored with simple examples, first from continuous waves in a semi-infinite medium and then within a guided wave structure.

## 2. Theory

The curl operator has a central role in vector calculus and in rotational waves including shear waves. In cartesian coordinates it is defined as:

$$\nabla \times F = \left( \frac{\partial F_z}{\partial y} - \frac{\partial F_y}{\partial z} \right) \hat{\boldsymbol{i}} + \left( \frac{\partial F_x}{\partial z} - \frac{\partial F_z}{\partial x} \right) \hat{\boldsymbol{j}} + \left( \frac{\partial F_y}{\partial x} - \frac{\partial F_x}{\partial y} \right) \hat{\boldsymbol{k}} = \begin{bmatrix} \dfrac{\partial F_z}{\partial y} - \dfrac{\partial F_y}{\partial z} \\[2mm] \dfrac{\partial F_x}{\partial z} - \dfrac{\partial F_z}{\partial x} \\[2mm] \dfrac{\partial F_y}{\partial x} - \dfrac{\partial F_x}{\partial y} \end{bmatrix}. \quad (1)$$

The curl operator is well defined for analytical expressions used in the later sections. For experimental results, finite differences and more sophisticated approaches can be used as an approximation given discrete samples of the vector displacement fields. The computational issues can lead to additional levels of complexity [18, 19] that are beyond the scope of this paper.

Directional filters can be applied to 2D and 3D spatial distributions, and with an additional dimension of time in transient applications. These can take a variety of forms but bear a general relation to the Fourier transform of the detected wave, and also the limitations of filtering given limited windows (limited support of filters). The Gabor directional filters are a classical example of practical filters that can discriminate and separate different orientations [20], albeit within the fundamental limitations on the ability to discriminate in both the image or spatial domain and the transform domain.

The well-known uncertainty relationship inherent to Fourier transform operations [21] is paramount in elastography because tissues and organs are inherently varying (grey vs. white matter in the brain, cortex vs. medulla in the kidney), and of limited size so limited support of directional filters is a practical issue, leading directly to constraints on discriminating between different wave directions for the purpose of analysis. These issues will be illustrated in the next sections.

## 3. Limitations Within a Semi-Infinite Medium

A canonical situation highly relevant to elastography is the simple sinusoidal excitation source on a semi-infinite medium. This was treated adeptly by Miller and Pursey [22] and also placed in context by Graff [1]. Three major categories of waves are produced by a small contact vibrating on the surface: longitudinal waves, shear waves, and surface or Rayleigh waves. These are illustrated in **Figure 1**.

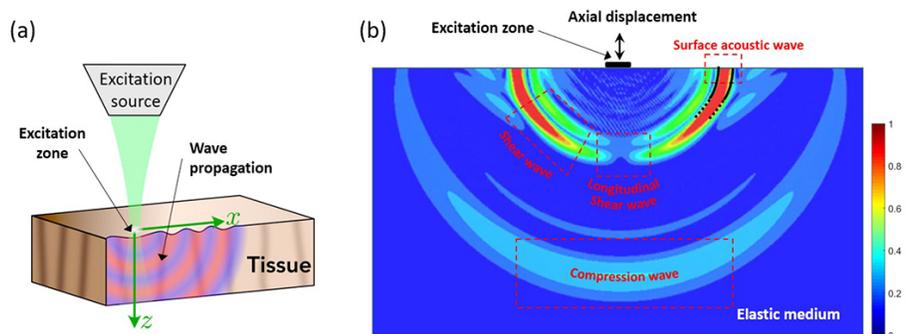

**Figure 1** Mechanical waves in optical coherence elastography. In **(a)**, axial motion at the surface of a tissue sample is produced by an excitation source for the generation of mechanical wave propagation. **(b)** Numerically simulated diagram depicting different mechanical wave branches generated when an axial harmonic load is applied at the surface of a semi-infinite elastic medium. Four waves are identified: surface acoustic wave (traveling along the surface), shear wave, compressional wave and longitudinal shear wave (both traveling towards increasing depth). Colormap represent normalized displacement magnitude in arbitrary units. [23]

### 3.1. The 2D Imaging Plane in Arbitrary Location

It can be simply noted from **Figure 1(a)** and **1(b)** that arbitrary imaging planes having an uncertain position with respect to the single source can lead to problems with interpretation. Consider an imaging plane parallel to the surface, such that the imaged region of interest (ROI) is oriented along the *x*-axis in **Figure 1** and extends out of page. If the ROI is positioned under the source, a "bulls eye" pattern of displacement emerges with the correct temporal frequency but a wavelength that appears anomalously large and bidirectional. In other cases with an arbitrary angle with respect to the surface, the unknown angle and vector projection of the propagating wave onto that plane leads to biased estimates towards larger wavelengths [24]. Neither curl nor directional filters can resolve this problem within a single 2D imaging plane, instead the exact positioning with respect to the source is required. This problem can be resolved by approaches that align the source and imaging plane: Echosens Fibroscan and acoustic radiation force systems (ARFI) are prominent examples of fixed configurations [25, 26]. Alternatively, examining a set of estimates across different angles in a freehand system has been proposed [24] within the context of the probability distribution of sampled wavelengths.

### 3.2. The 2D Imaging Plane With Several Shear Wave Sources

This is a practical situation, illustrated with a pair of sources and the superposition of the Miller and Pursey [22] solution (**Figure 1**) in order to generate a pair of waves that superimpose with some angular alignment [27, 28]. **Figure 2** describes a far field simplification of the middle region between two small external sources vibrating normal to the surface of a semi-infinite medium. A pair of plane waves are represented having identical amplitude and angle with respect to the depth ($x$-axis). Along the midline, the shear vectors add to create an $x$-directed displacement. To an imaging system sensitive to the $x$-axis, this appears similar to a longitudinal shear wave but with an elongated wavelength (higher shear speed than the nominal shear wave velocity of $c_s = \sqrt{G/\rho}$, where $G$ is the shear modulus and $\rho$ is the density of the elastic material). The error is proportional to $1/\cos(\theta)$, where $\theta$ is the angle with respect to the $x$-axis in **Figure 2**, and this error is not readily resolved for the case of the $x$-sensitive 2D imaging system.

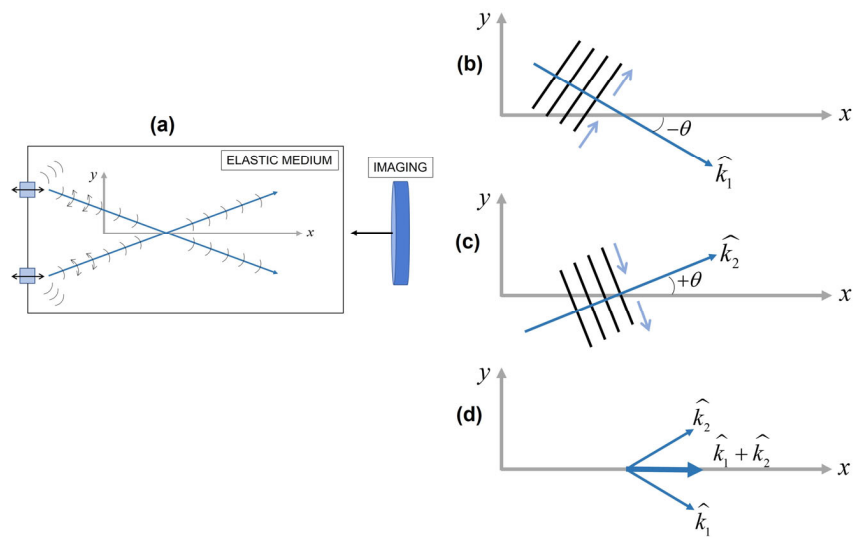

**Figure 2.** Two external sources are applied to a semi-infinite medium (a) Approximating the far field of the Miller-Pursey solution (**Figure 1**), we examine a pair of polarized shear waves propagating in the $\widehat{k_1}$ and $\widehat{k_2}$ direction with transverse particle displacements ((b) and (c), respectively). The superposition of these along the center line produces an $\hat{x}$-directed displacement with an apparent wavelength longer than the true shear wave wavelength.

Directional filters applied to the case of the superimposed pair of waves shown in **Figures 2 and 3** can fail to enable an accurate assessment of the underlying components with the correct wavelength. Directional filters can take many forms in two-dimensional (2D) or 3D spatial transforms or mixed spatial temporal filters. Many bear some similarity to the classical Gabor directional filters that have been widely used in imaging applications [20] along with the inherent limits on resolution between the spatial and transform domains [21].

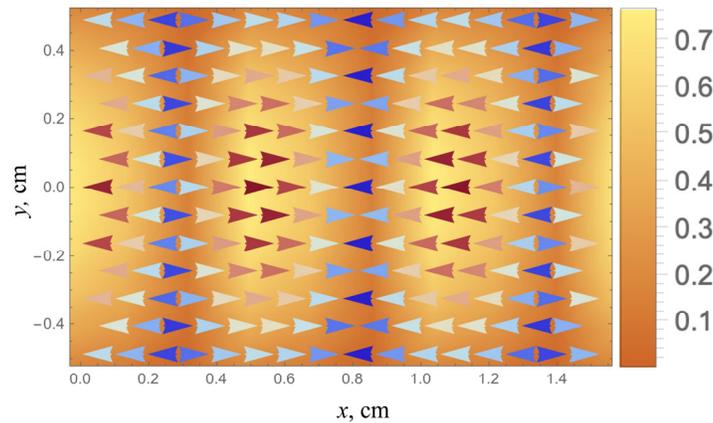

**Figure 3.** Vector and amplitude representation of the *x*-component displacements within a ROI taken from the configuration of **Figure 2**. The vertical axis $y = 0$ in the middle of this figure is the centerline between two small external sources and with angle theta set to $\pi/8$. The pattern appears to resemble a longitudinal shear wave pattern and with a wavelength in the *x*-direction that is biased upward from the shear wavelength.

To illustrate the limitations, we examine the 2D spatial Fourier transform of the pair of plane waves produced by two external sources as illustrated in **Figure 2** and with the angle $\theta$ taken as $\pi/16$ radians. The practical issue here is that we will necessarily have a finite window on the function, so instead of obtaining $\Im\{U_x(x,y)\}$, we obtain $\Im\{W(x,y)U_x(x,y)\}$, where $\Im\{\ \}$ is the 2D spatial transform operator, $U_x(x,y)$ is the measured displacements in the *x* direction, and $W(x,y)$ is the spatial window applied, where small windows are typically preferred for spatial resolution. **Figure 4(a)** shows the magnitude of the transform utilizing a square window of 6 cm on edge ($6\lambda$ in this example), close enough to the ideal transform where we can distinguish two different waves and can reasonably filter these with finite impulse response directional filters of the same spatial support. However, $6\lambda$ windows are very unlikely to be achieved in many elastography studies, for example at 50 Hz shear wave frequency and assuming $c_s = 1$ m/s in soft tissue, the wavelength is 2 cm, so a $6\lambda$ window of 12 cm on a side will average over many substructures of interest within a large organ like the brain or kidney. More realistically, **Figure 4(b)** shows the transform magnitude with a $2\lambda$ (2 cm) square window. Here, the transform indicates a predominantly *x*-directed wave, commensurate with the appearance of **Figure 3** and with error and uncertainty in the nominal wavelength proportional in this simple example to $1/\cos(\theta)$.

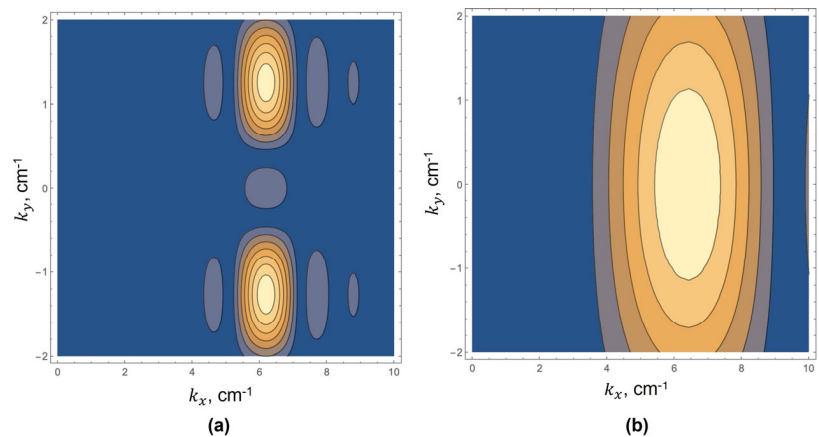

**Figure 4** Magnitude plots in contour form of the right half ($+k_x$) of the spatial Fourier transform of the pair of waves described in **Figure 2**, however with practical limits on the rectangular window limiting the extent of the ROI. In (a), a rectangular window of $6\lambda$ is applied, and the resulting transform can resolve two waves separated by an angle $\phi$. However, in (b), a rectangular window of only $2\lambda$ is applied and the resulting uncertainty in the spatial frequency domain blurs the peaks leading to a biased estimate of $\hat{k}$. This fundamental window effect also pertains to the limited support of directional filters.

## 4. Limitations Within Guided Wave Structures

Things get more complicated when a structure such as a tendon or cornea can modify the phase velocity by guiding a wave within its boundaries. The exact solution will depend on boundary conditions, shape, size, and material properties, but the key message here is that a wide range of different phase velocities are possible at a single frequency, and no one of these phase velocities (at a single frequency) is likely to match the phase velocity of shear waves, $c_s$, expected within an infinite medium. The theory behind these cases was magnificently illustrated in the 1870s through the early 1900s with landmark papers from Pochhammer [29], Rayleigh [30], Love [31], and Lamb [32, 33].

We will follow Lamb's 1917 derivation because this spurred far-reaching research for over a century of new applications, eventually arriving in elastography where structures such as biofilms [34], corneas [23, 35], and arterial walls [36] are under investigation. An illustration of the two possible modes, symmetric and antisymmetric, propagating in a thin structure with parallel sides is shown in **Figure 5**.

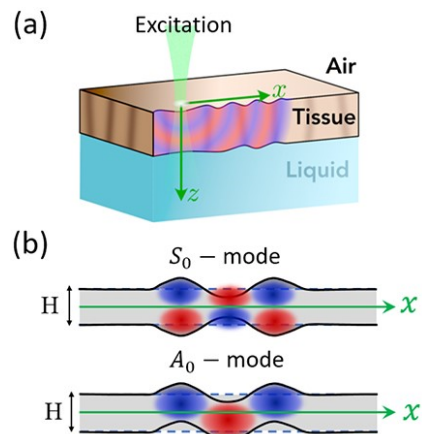

**Figure 5** Propagation of Lamb waves in tissues. **(a)** Thin-plate type tissue (interfacing air at the top and liquid fluid at the bottom) being locally excited with axial motion at the top surface. **(b)** Lamb waves are generated and guided by the thin-plate in the quasi-symmetric ($S_0$) and quasi-antisymmetric ($A_0$) zero-order modes. Red and blue fields represent positive and negative displacement, respectively. [23]

A significant issue for elastography of structures that can guide waves is that the modes allow for propagation in the long axis of periodic displacements that have a phase velocity below, above, or much higher than the nominal shear wave velocity of $c_s = \sqrt{G/\rho}$, where $G$ is the shear modulus and $\rho$ is the density of the elastic material. An excitation at one frequency can, in theory and in practice, excite more than one mode of propagation, each with a separate phase velocity, and none matching the nominal shear wave velocity [2, 37, 38]. A numerical example following the derivations of Lamb [33] is given in the **Appendix** for the purpose of illustration. However, the main question before us is: in the case of a guided wave structure, will either the curl operator or some set of directional filters provide us a direct estimate of spatial frequency directly and solely related to $c_s$? The answer is no, at least not without rare and special conditions. The form of the solutions from Pochhammer, Rayleigh, Love, and Lamb are all separable with a phase velocity in the long axis $z$ given as $\exp[I\xi z]$, where $\xi$ is the wavenumber derived from a complicated interaction of factors including boundary conditions, geometry, and the longitudinal wave speed, and $I$ is the imaginary unit. The component directions of the propagations are separable along the long axis and in the lateral directions respectively, with no need for filters to separate them, however no spatial wavenumber will reduce to the nominal $k_s = \omega/c_s$ except asymptotically at relatively high frequencies. There is, however, a way to more fully characterize the structure under examination: given *a priori* knowledge of the shape and size of the structure and measurements of a particular mode across different frequencies, an inverse solution can be generated for the unknown material properties [39-41]. In more general terms, the measured dispersion of the phase velocity can be input to appropriate models to estimate the unknown shear modulus of a waveguide.

For the case of Lamb waves, we can show that the curl operator does provide a periodic pattern that can be further analyzed to estimate the shear modulus. An example is given in **Figure 6** of the symmetric mode of the soft waveguide at 100 Hz described more fully in the **Appendix**. The shear wave speed in this example is $c_s = 1$ m/s and we would like to estimate that from the observed displacements. However, we find repeating patterns in the symmetric mode and in the result of the curl operator that are periodic along the long axis with the apparent wavelength consistent with twice the shear velocity, where cycles repeat every 2 cm instead of every 1 cm in the long axis. However, it can be shown that by further applying the Laplacian operator in two dimensions, then using this within a Helmholtz direct inversion formula, the result is capable of providing an estimate of the shear modulus, as described in more detail in the **Appendix**.

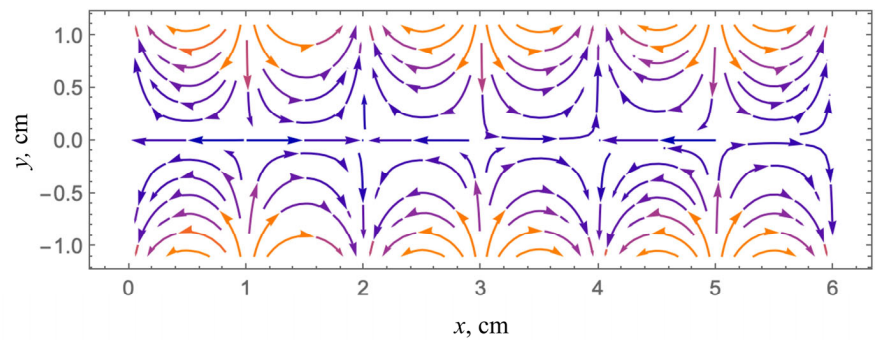

**Figure 6** Vector streamline diagram of 6 cm segment of an infinite plate in the horizontal direction and out-of-page direction, with propagating symmetric Lamb waves at 100 Hz. Periodic repeats are seen every 2 cm, however a shear wave of this material in an infinite medium would have a wavelength of 1 cm. The curl operator and directional filters will not provide a direct 1 cm wavelength, further steps are required.

Another important issue with the use of curl in this example, and more generally in elastography, is that estimates of displacements in tissues tend to be noisy due to the small, micronscale displacements commonly utilized across all the imaging modalities, along with the need for speed in data acquisition (limited averaging across repetitions) and the reality of physiological motions and electronic noise. Taking the curl's spatial derivatives using finite difference approximations will notoriously amplify noise. In this example one must estimate the curl and then the Laplacian in order to complete an estimation of the shear wave speed, essentially three successive orders of spatial derivatives which amplify noise. This would be followed by a quotient of terms for the direct Helmholtz equations [42, 43]. Thus, more sophisticated noise reduction and regularization strategies may be required [44, 45].

## 5. Discussion

In optical coherence tomography, MRI, and ultrasound imaging systems, one of the fastest means of obtaining information for elastography is to obtain a single component of displacement or velocity in a 2D imaging plane. For common situations where external sources are providing continuous shear waves, simple configurations are shown to be resistant to improved analysis by the use of curl or directional filters. The curl operator requires volumetric information which is not available in simple 2D imaging ROIs. Directional filters' ability to separate and discriminate between different waves is limited by the support of the filters and the size of the ROI, such that practical configurations resembling a pair of Miller-Pursey sources can produce wavefronts that are easily misinterpreted and not resolvable by directional filters.

One means of mitigating these difficulties is to design the source and imaging plane in a fixed geometry, for example in Echosens and ARFI systems where the excitation and imaging planes are preconfigured [25, 26].

Another means of mitigating the difficulties is to establish a 3D set of shear waves across multiple directions, producing the reverberant shear wave field. In that case, the imaging system ROI can have an arbitrary configuration with respect to the organ of interest, and the estimators take advantage of the simple limiting mathematics of the expected ensemble of waves in 3D [46-48]. This enables freehand orientation of the imaging plane, as is common with hand-held imaging probes, in addition to preset 3D scan geometries.

A limitation of this paper is that for both the curl operator and the directional filters, their ability to improve estimates of the shear modulus will diminish as a function of specific parameters, including the noise level in the displacement estimates and the size of the imaging ROI relative to the wavelength. Further work is required to characterize these error curves and trends for specific implementations.

**Funding:** This research was funded by the National Institutes of Health, grant number R01DK126833.

**Data Availability Statement:** The data are contained within the paper, equations, and Appendix.

**Acknowledgments:** The author is grateful for insightful comments and perspectives from Professor Robert Rohling, Professor Samuel Patz, and Professor Fernando Zvietcovich.

**Conflicts of Interest:** The author declares no conflict of interest.

## Appendix

Following Lamb [33], we calculate the waves within a long parallel plate, with stress-free boundary conditions on the top and bottom surfaces, infinite in the $x$ and $z$ directions, with waves propagating in the $x$ direction and all $z$ dependence and derivatives equal to zero. We quote from Lamb [33] directly for the following section, and use his original notation where $\sigma$ is radial frequency, $\lambda$ and $\mu$ are the Lamé constants, $\rho$ is density, $h$ is the wavenumber associated with longitudinal waves in an infinite medium, $k$ is the wavenumber associated with shear waves in an infinite medium, and $u$ and $v$ are the wave displacements in the $x$ and $y$ directions, respectively. Furthermore, $x$ is the direction of the long axis of the plate and the thickness in the $y$ axis is defined as $2f$, A and B are amplitudes of the wave functions (constants), $\phi$ and $\psi$ are the wavefunctions associated with longitudinal and shear wave propagation, respectively, and $\alpha$ and $\beta$ are additional wavenumbers. The wavelength describing the resulting periodicity in the long axis is $\lambda'$. Lamb's derivation proceeds as follows:

"Then we assume a time factor $e^{i\sigma t}$ (omitted in the sequel), and write:

$$h^2 = \frac{\rho\sigma^2}{\lambda + 2\mu}, \qquad k^2 = \frac{\rho\sigma^2}{\mu}, \tag{2}$$

so that

$$\left(\nabla^2 + h^2\right)\phi = 0, \qquad \left(\nabla^2 + k^2\right)\psi = 0. \tag{3}$$

We further assume, for the present purpose, periodicity with respect to $x$. This is most conveniently done by means of a factor of $e^{i\xi x}$, the wavelength being accordingly:

$$\lambda' = \frac{2\pi}{\xi}. \tag{4}$$

Writing

$$\alpha^2 = \xi^2 - h^2, \qquad \beta^2 = \xi^2 - k^2, \tag{5}$$

we have

$$\frac{\partial^2 \phi}{\partial y^2} = \alpha^2 \phi, \qquad \frac{\partial^2 \psi}{\partial y^2} = \beta^2 \psi. \tag{6}$$

The solutions for the internal stresses are:

$$\frac{p_{yy}}{\mu} = \left\{ A\left(\xi^2 + \beta^2\right)\cos\alpha f - B2i\xi\beta\cos\beta f \right\}e^{i\xi x},$$

$$\frac{p_{yy}}{\mu} = \pm\left\{ A2i\xi\alpha\sin\alpha f + B\left(\xi^2 + \beta^2\right)\sin\beta f \right\}e^{i\xi x}$$

(7)

Equating these…" to the boundary conditions provide the allowable interrelationships between the constants. Equations (5) and (6) are more profound than first meets the eye; they effectively change the wavenumbers from simple dependence on shear and longitudinal wave speeds to a more complicated set of wave speeds related to $\alpha$, $\beta$, and $\xi$, which in turn are related to a number of material factors and boundary conditions. Finally, Lamb also states that when the motion is symmetrical with respect to the plane $y = 0$ we assume, in accordance with eqn (6):

$$\phi = A\cos\alpha y e^{i\xi x}, \qquad \psi = B\sinh\beta y e^{i\xi x},$$

(8)

$$u = \frac{\partial\phi}{\partial x} + \frac{\partial\psi}{\partial y}, \qquad v = \frac{\partial\phi}{\partial y} - \frac{\partial\psi}{\partial x}.$$

(9)

The equations for displacement in the $x$ and $y$ directions are then given as:

$$u = I A\xi e^{I\xi x}\cos\alpha y + B\beta e^{I\xi x}\cos\beta y,$$

$$v = A\alpha e^{i\xi x}\sin\alpha y - IB\xi e^{I\xi x}\sin\beta y.$$

(10)

Consistent with the long wavelength approximations (Lamb's eqns (10-15) and related discussion), we have chosen an example where a soft biological material ($G = 1000$ Pa, $\omega = 2\pi \times 100$ s⁻¹, $c_{ph} = 1$ m/s, $\lambda = 1$ cm) of thickness 2 cm in the $y$ direction and with a wave propagating in the $x$ direction. The parameters for eqn (10) are also taken as consistent with Lamb's remarks for the long wavelength approximations: $A = B = 1$, $\omega = 2 \times \pi \times 100$ s⁻¹, $\alpha = 99 \times \pi$ m⁻¹, $\xi = 100 \times \pi$ m⁻¹, and $\beta = I \times 100 \times \sqrt{3}\pi$ m⁻¹. The resulting vector field and its curl are periodic in the $x$ direction with an evident wavelength of 2 cm, instead of 1 cm as would be the case for a shear wave in an infinite medium of this material. The vector plot of a symmetric mode at 100 Hz is given in **Figure 6**.

Although the wavelength and phase velocity in the $x$ direction are a factor of 2 higher than in an infinite medium, the curl of this vector field is found by substituting eqn (10) into eqn (1) with the $z$ derivatives assumed to be zero. The result is entirely in the $z$ direction (out of page) and is of the form $Be^{i\xi x}(-\beta^2 + \xi^2)\sinh\beta y$. The Laplacian of this has the form $-Be^{I\xi x}(\beta^2 - \xi^2)^2\sinh\beta y$. Thus, a division of the two representing a direct inversion of the Helmholtz equation produces the term $(\beta^2 - \xi^2)$, which according to eqn (5) is equal to $k^2$, the wavenumber associated with the shear modulus. It can be shown that similar results are found for the antisymmetric case. This implies that within the guided wave structures having highly anomalous phase velocities in the long axis, the curl operator only reproduces the Lamb wavelength (not the nominal shear wavelength), but this result can be used as a first step towards isolating the shear wavenumber $k$, and therefore $c_s$. Lamb referred to similarities between his approach and with Pochhammer's solutions for cylinders, so this is likely to be the case across structures of different cross-sectional shapes, although further work is required to confirm that hypothesis.

## References

1.  Graff, K. F. *Wave motion in elastic solids*. Oxford: Clarendon Press, 1975, 356.

2.  Baghani, A., S. Salcudean and R. Rohling. "Theoretical limitations of the elastic wave equation inversion for tissue elastography." *J Acoust Soc Am* 126 (2009): 1541. 10.1121/1.3180495. https://www.ncbi.nlm.nih.gov/pubmed/19739767.

3.    Sinkus, R., M. Tanter, T. Xydeas, S. Catheline, J. Bercoff and M. Fink. "Viscoelastic shear properties of in vivo breast lesions measured by mr elastography." *Magn Reson Imaging* 23 (2005): 159-65. S0730-725X(05)00039-1 [pii]

10.1016/j.mri.2004.11.060. http://www.ncbi.nlm.nih.gov/pubmed/15833607.

4.    Kwon, O. I., C. Park, H. S. Nam, E. J. Woo, J. K. Seo, K. J. Glaser, A. Manduca and R. L. Ehman. "Shear modulus decomposition algorithm in magnetic resonance elastography." *IEEE Trans Med Imaging* 28 (2009): 1526-33. 10.1109/tmi.2009.2019823.

5.    Honarvar, M., R. Sahebjavaher, R. Sinkus, R. Rohling and S. E. Salcudean. "Curl-based finite element reconstruction of the shear modulus without assuming local homogeneity: Time harmonic case." *IEEE Trans Med Imaging* 32 (2013): 2189-99. 10.1109/tmi.2013.2276060.

6.    Hirsch, S., J. Guo, R. Reiter, E. Schott, C. Büning, R. Somasundaram, J. Braun, I. Sack and T. J. Kroencke. "Towards compression-sensitive magnetic resonance elastography of the liver: Sensitivity of harmonic volumetric strain to portal hypertension." *J Magn Reson Imaging* 39 (2014): 298-306. 10.1002/jmri.24165.

7.    Manduca, A., P. J. Bayly, R. L. Ehman, A. Kolipaka, T. J. Royston, I. Sack, R. Sinkus and B. E. Van Beers. "Mr elastography: Principles, guidelines, and terminology." *Magn Reson Med* 85 (2021): 2377-90. 10.1002/mrm.28627.

8.    Baghani, A., R. Zahiri Azar, S. Salcudean and R. Rohling. "A curl-based approach to ultrasound elastography." Presented at ASME 2010 International Mechanical Engineering Congress and Exposition, 2010. 2: Biomedical and Biotechnology Engineering, 865-67. 10.1115/imece2010-39180.

9.    Hashemi, H. S., S. E. Salcudean and R. N. Rohling. *Ultrafast ultrasound imaging for 3d shear wave absolute vibro-elastography*. 2022, arXiv:2203.13949.

10.   Manduca, A., D. S. Lake, S. A. Kruse and R. L. Ehman. "Spatio-temporal directional filtering for improved inversion of mr elastography images." *Med Image Anal* 7 (2003): 465-73. http://www.ncbi.nlm.nih.gov/pubmed/14561551.

11.   Kruse, S. A., G. H. Rose, K. J. Glaser, A. Manduca, J. P. Felmlee, C. R. Jack, Jr. and R. L. Ehman. "Magnetic resonance elastography of the brain." *Neuroimage* 39 (2008): 231-7. S1053-8119(07)00716-1 [pii]

10.1016/j.neuroimage.2007.08.030. http://www.ncbi.nlm.nih.gov/pubmed/17913514.

12.   Deffieux, T., J. L. Gennisson, J. Bercoff and M. Tanter. "On the effects of reflected waves in transient shear wave elastography." *IEEE Trans Ultrason Ferroelectr Freq Control* 58 (2011): 2032-35. 10.1109/TUFFC.2011.2052.

13.   Song, P., H. Zhao, A. Manduca, M. W. Urban, J. F. Greenleaf and S. Chen. "Comb-push ultrasound shear elastography (cuse): A novel method for two-dimensional shear elasticity imaging of soft tissues." *IEEE Trans Med Imaging* 31 (2012): 1821-32. 10.1109/TMI.2012.2205586.

14.   Zhao, H., P. Song, D. D. Meixner, R. R. Kinnick, M. R. Callstrom, W. Sanchez, M. W. Urban, A. Manduca, J. F. Greenleaf and S. Chen. "External vibration multi-directional ultrasound shearwave elastography (evmuse): Application in liver fibrosis staging." *IEEE Trans Med Imaging* 33 (2014): 2140-8. 10.1109/tmi.2014.2332542.

15.   Song, S., N. M. Le, Z. Huang, T. Shen and R. K. Wang. "Quantitative shear-wave optical coherence elastography with a programmable phased array ultrasound as the wave source." *Opt Lett* 40 (2015): 5007-10. 10.1364/ol.40.005007.

16.   Tzschatzsch, H., S. Ipek-Ugay, M. N. Trong, J. Guo, J. Eggers, E. Gentz, T. Fischer, M. Schultz, J. Braun and I. Sack. "Multifrequency time-harmonic elastography for the measurement of liver viscoelasticity in large tissue windows." *Ultrasound Med Biol* 41 (2015): 724-33. 10.1016/j.ultrasmedbio.2014.11.009. http://www.ncbi.nlm.nih.gov/pubmed/25638319.

17.   Khodayi-Mehr, R., M. W. Urban, M. M. Zavlanos and W. Aquino. "Plane wave elastography: A frequency-domain ultrasound shear wave elastography approach." *Phys Med Biol* 66 (2021): 10.1088/1361-6560/ac01b8.

18.   Bochev, P. B., C. J. Garasi, J. J. Hu, A. C. Robinson and R. S. Tuminaro. "An improved algebraic multigrid method for solving maxwell's squations." *SIAM J Sci Comput* 25 (2003): 623-42. 10.1137/s1064827502407706. https://epubs.siam.org/doi/abs/10.1137/S1064827502407706.

19.  Hochbruck, M. and T. Pažur. "Implicit runge--kutta methods and discontinuous galerkin discretizations for linear maxwell's equations." *SIAM J Numer Anal* 53 (2015): 485-507. 10.1137/130944114. https://epubs.siam.org/doi/abs/10.1137/130944114.

20.  Daugman, J. G. "Uncertainty relation for resolution in space, spatial frequency, and orientation optimized by two-dimensional visual cortical filters." *J Opt Soc Am A* 2 (1985): 1160-9. 10.1364/josaa.2.001160.

21.  Bracewell, R. N. *Two-dimensional imaging*. Englewood Cliffs, N.J.: Prentice Hall, 1995,

22.  Miller, G. F. and H. Pursey. "The field and radiation impedance of mechanical radiators on the free surface of semi-infinite isotropic solid." *Proc R Soc Lond A* 223 (1954): 521-41. 10.1098/rspa.1954.0134. http://rspa.royalsocietypublishing.org/content/223/1155/521.full.pdf+html.

23.  Zvietcovich, F. and K. V. Larin. "Wave-based optical coherence elastography: The 10-year perspective." *Prog Biomed Eng (Bristol)* 4 (2022): 012007. 10.1088/2516-1091/ac4512.

24.  Tzschatzsch, H., S. Ipek-Ugay, J. Guo, K. J. Streitberger, E. Gentz, T. Fischer, R. Klaua, M. Schultz, J. Braun and I. Sack. "In vivo time-harmonic multifrequency elastography of the human liver." *Phys Med Biol* 59 (2014): 1641-54. 10.1088/0031-9155/59/7/1641. http://www.ncbi.nlm.nih.gov/pubmed/24614751.

25.  Sarvazyan, A. P., O. V. Rudenko, S. D. Swanson, J. B. Fowlkes and S. Y. Emelianov. "Shear wave elasticity imaging: A new ultrasonic technology of medical diagnostics." *Ultrasound Med Biol* 24 (1998): 1419-35. S0301-5629(98)00110-0 [pii]. http://www.ncbi.nlm.nih.gov/pubmed/10385964.

26.  Sandrin, L., B. Fourquet, J. M. Hasquenoph, S. Yon, C. Fournier, F. Mal, C. Christidis, M. Ziol, B. Poulet, F. Kazemi, *et al.* "Transient elastography: A new noninvasive method for assessment of hepatic fibrosis." *Ultrasound Med Biol* 29 (2003): 1705-13. S0301562903010718 [pii]. http://www.ncbi.nlm.nih.gov/pubmed/14698338.

27.  Wu, Z., L. S. Taylor, D. J. Rubens and K. J. Parker. "Shear wave focusing for three-dimensional sonoelastography." *J Acoust Soc Am* 111 (2002): 439-46. http://www.ncbi.nlm.nih.gov/pubmed/11831818.

28.  Partin, A., Z. Hah, C. T. Barry, D. J. Rubens and K. J. Parker. "Elasticity estimates from images of crawling waves generated by miniature surface sources." *Ultrasound Med Biol* 40 (2014): 685-94. http://dx.doi.org/10.1016/j.ultrasmedbio.2013.05.019. http://www.sciencedirect.com/science/article/pii/S030156291300793X.

29.  Pochhammer, L. "Ueber die fortpflanzungsgeschwindigkeiten kleiner schwingungen in einem unbegrenzten isotropen kreiscylinder." *Journal für die reine und angewandte Mathematik* 1876 (1876): 324-36. doi:10.1515/crll.1876.81.324. https://doi.org/10.1515/crll.1876.81.324.

30.  Rayleigh, L. "On the free vibrations of an infinite plate of homogeneous isotropic elastic matter." *Proceedings of the London Mathematical Society* s1-20 (1888): 225-37. https://doi.org/10.1112/plms/s1-20.1.225. https://londmathsoc.onlinelibrary.wiley.com/doi/abs/10.1112/plms/s1-20.1.225.

31.  Love, A. E. H. *A treatise on the mathematical theory of elasticity*. 2d. Cambridge: University Press, 1906,

32.  Lamb, H. "On the flexure of an elastic plate." *Proceedings of the London Mathematical Society* s1-21 (1889): 70-91. 10.1112/plms/s1-21.1.70. https://doi.org/10.1112/plms/s1-21.1.70.

33.  Lamb, H. "On waves in an elastic plate." *Proceedings of the Royal Society of London. Series A, Containing Papers of a Mathematical and Physical Character* 93 (1917): 114-28. doi:10.1098/rspa.1917.0008. https://royalsocietypublishing.org/doi/abs/10.1098/rspa.1917.0008.

34.  Mercado, K. P., J. Langdon, M. Helguera, S. A. McAleavey, D. C. Hocking and D. Dalecki. "Scholte wave generation during single tracking location shear wave elasticity imaging of engineered tissues." *J Acoust Soc Am* 138 (2015): EL138-EL44. doi:http://dx.doi.org/10.1121/1.4927633. http://scitation.aip.org/content/asa/journal/jasa/138/2/10.1121/1.4927633.

35.  Kirby, M. A., I. Pelivanov, S. Song, Ł. Ambrozinski, S. J. Yoon, L. Gao, D. Li, T. T. Shen, R. K. Wang and M. O'Donnell. "Optical coherence elastography in ophthalmology." *J Biomed Opt* 22 (2017): 1-28. 10.1117/1.Jbo.22.12.121720.

36. Couade, M., M. Pernot, C. Prada, E. Messas, J. Emmerich, P. Bruneval, A. Criton, M. Fink and M. Tanter. "Quantitative assessment of arterial wall biomechanical properties using shear wave imaging." *Ultrasound Med Biol* 36 (2010): 1662-76. 10.1016/j.ultrasmedbio.2010.07.004.

37. di Novi, R. A. "Theory of lamb waves." Presented at Symposium on Physics and Nondestructive Testing: Held at Argonne National Laboratory October 4 and 5, 1960, 1960. Argonne National Laboratory, Metallurgy Division, 6346, 72.

38. Pelivanov, I., L. Gao, J. Pitre, M. Kirby, S. Song, D. Li, T. Shen, R. Wang and M. O'Donnell. "Does group velocity always reflect elastic modulus in shear wave elastography?" *J Biomedl Opt* 24 (2019): 076003. 10.1117/1.JBO.24.7.076003. https://doi.org/10.1117/1.JBO.24.7.076003.

39. Pagneux, V. and A. Maurel. "Determination of lamb mode eigenvalues." *J Acoust Soc Am* 110 (2001): 1307-14. 10.1121/1.1391248.

40. Han, Z., S. R. Aglyamov, J. Li, M. Singh, S. Wang, S. Vantipalli, C. Wu, C. H. Liu, M. D. Twa and K. V. Larin. "Quantitative assessment of corneal viscoelasticity using optical coherence elastography and a modified rayleigh-lamb equation." *J Biomed Opt* 20 (2015): 20501. 10.1117/1.Jbo.20.2.020501.

41. Sun, M. G., T. Son, J. Crutison, V. Guaiquil, S. Lin, L. Nammari, D. Klatt, X. Yao, M. I. Rosenblatt and T. J. Royston. "Optical coherence elastography for assessing the influence of intraocular pressure on elastic wave dispersion in the cornea." *J Mech Behav Biomed Mater* 128 (2022): 105100. 10.1016/j.jmbbm.2022.105100.

42. Manduca, A., T. E. Oliphant, M. A. Dresner, J. L. Mahowald, S. A. Kruse, E. Amromin, J. P. Felmlee, J. F. Greenleaf and R. L. Ehman. "Magnetic resonance elastography: Non-invasive mapping of tissue elasticity." *Med Image Anal* 5 (2001): 237-54. S1361841500000396 [pii]. http://www.ncbi.nlm.nih.gov/pubmed/11731304.

43. Doyley, M. M. "Model-based elastography: A survey of approaches to the inverse elasticity problem." *Phys Med Biol* 57 (2012): R35-R73. 10.1088/0031-9155/57/3/R35. http://www.ncbi.nlm.nih.gov/pubmed/22222839.

44. de Felice, G., F. S. Marra and G. C. Rufolo. "Regularized solutions for the discrete forms of the div–curl problem in cfd." *Computing and Visualization in Science* 4 (2002): 175-82. 10.1007/s007910100069. https://doi.org/10.1007/s007910100069.

45. Dong, H. and G. D. Egbert. "Divergence-free solutions to electromagnetic forward and adjoint problems: A regularization approach." *Geophys J Int* 216 (2018): 906-18. 10.1093/gji/ggy462. https://doi.org/10.1093/gji/ggy462.

46. Aleman-Castañeda, L. A., F. Zvietcovich and K. J. Parker. "Reverberant elastography for the elastic characterization of anisotropic tissues." *Ieee Journal of Selected Topics in Quantum Electronics* 27 (2021): 1-12. 10.1109/JSTQE.2021.3069098.

47. Ge, G. R., W. Song, M. Nedergaard, J. P. Rolland and K. J. Parker. "Theory of sleep/wake cycles affecting brain elastography." *Phys Med Biol* 67 (2022): 225013. 10.1088/1361-6560/ac9e40. https://dx.doi.org/10.1088/1361-6560/ac9e40.

48. Kabir, I. E., D. A. Caban-Rivera, J. Ormachea, K. J. Parker, C. L. Johnson and M. M. Doyley. "Reverberant magnetic resonance elastographic imaging using a single mechanical driver." *Phys Med Biol* 68 (2023): 10.1088/1361-6560/acbbb7.